\documentclass[pra,twocolumn,superscriptaddress,amsmath]{revtex4}
\usepackage{graphicx}
\usepackage{color}

\begin{document}

\def\ket#1{|#1\rangle}
\def\bra#1{\langle#1|}
\def\av#1{\langle#1\rangle}
\def\myarrow{\mathop{\longrightarrow}}

\title{Mean-field treatment of the damping of the oscillations of a 1D Bose  gas in an
optical lattice
}

\author{Julio Gea-Banacloche}
\affiliation{Department of Physics, University of Arkansas, Fayetteville, AR 72701}
\email[]{jgeabana@uark.edu}
\author{Ana Mar\'\i a Rey}
\affiliation{National Institute of Standards and Technology, Gaithersburg, MD 20899}
\author{Guido Pupillo}
\affiliation{National Institute of Standards and Technology, Gaithersburg, MD 20899}
\author{Carl J. Williams}
\affiliation{National Institute of Standards and Technology, Gaithersburg, MD 20899}
\author{Charles W. Clark}
\affiliation{National Institute of Standards and Technology, Gaithersburg, MD 20899}

\date{\today}

\begin{abstract}
We present a theoretical treatment of the surprisingly large damping observed recently in
 one-dimensional Bose-Einstein atomic condensates in optical lattices.  We show that time-dependent Hartree-Fock-Bogoliubov (HFB) calculations can describe qualitatively the main features of the damping observed over a range of lattice depths.  We also derive a formula of the fluctuation-dissipation type for the damping, based on a picture in which the coherent motion of the condensate atoms is disrupted as they try to flow through the random local potential created by the irregular motion of noncondensate atoms.  We expect this irregular motion to result from the well-known dynamical instability exhibited by the mean-field theory for these systems.  When parameters for the characteristic strength and correlation times of the fluctuations, obtained from the HFB calculations, are substituted in the damping formula, we find very good agreement with the experimentally-observed damping, as long as the lattice is shallow enough for the fraction of atoms in the Mott insulator phase to be negligible.  We also include, for completeness, the results of other calculations based on the Gutzwiller ansatz, which appear to work better for the deeper lattices.  
\end{abstract}
\maketitle

\section{Introduction}

The transport properties of atomic Bose-Einstein condensates have recently been the subject of much interest.
 In a pure harmonic trap, the dipole mode of the motion---where the cloud of atoms oscillates back and forth without
  altering its shape---is known to be stable.  On the other hand, if an optical lattice is used to create a
one-dimensional array of potential wells and barriers, one may
find, even in a single-particle picture, a damping of the
oscillations due to the non-quadratic nature of the resulting
dispersion relation \cite{Ruuska, Pezze,Ott,Rigol,Rigol3,ana2}.  When interactions
between atoms are included, at the mean-field level, one finds dynamical instabilities
\cite{wu} that may result in a very large damping \cite{burger,
Smerzi}. All these effects are, however, only expected to be
substantial  when the quasimomentum of the cloud of atoms is
sufficiently large (typically, of the order of $\pi\hbar/\lambda$,
where $\lambda/2$ is the lattice spacing).

In recent experiments with ${}^{87}$Rb atoms \cite{trey,esslinger},
confined to move in one-dimensional ``tubes,''
 a surprisingly large damping of the dipole mode was observed, for very weak optical lattices and very
 small cloud displacements.  We note that no (or very little) damping was observed for the same system
 in the absence of the tight transverse confinement \cite{Morsch,Cataliotti}.  In the experiments \cite{trey}, the oscillation frequency in the harmonic trap,
$\omega_0/2\pi$ was about 60 Hz, whereas the photon recoil energy
$E_R=h^2/(2m\lambda^2)$ corresponded to a frequency $E_R/h=3.47$
kHz.  Under these conditions, for a shallow lattice, the maximum
displacement of the condensate in the experiment (7 to 8 lattice
sites) should not result in a momentum larger than about
$0.1(\pi\hbar/\lambda)$, which is well within the quadratic part
of the lattice dispersion curve. Likewise, the quasimomentum
spread arising from the finite size of the cloud itself was also
quite small (of the order of $2 \pi\hbar/13 \lambda$, since the
Thomas-Fermi radius of the cloud is about $13 \lambda$).

Since these results were first presented (and, in some cases, predating them), a number of theoretical treatments have been put forward that, directly or indirectly, address various relevant aspects of the underlying dynamics, from different perspectives.  It has been shown, for instance \cite{polkovnikov,altman,polkovnikov2}, that the momentum cutoff for the dynamical instability may be substantially lowered for commensurate lattices, and, probably more relevant for the experimental situation, that the boundary between regular and irregular motion becomes ``smeared out'' due to quantum fluctuations.  Numerical calculations based on a truncated Wigner representation \cite{ruostekoski} have also shown that the fraction of atoms with momenta in the unstable region can indeed cause damping of the center of mass motion of the whole system.  As we shall show below, this fraction is, in fact, a non-negligible number, for the experimental parameters, even for relatively shallow lattices, because of the large depletion caused by the very tight transverse confinement.

In a recent series of papers \cite{ana2,guido}, several of us have characterized the damping mechanisms that may dominate,  for these systems, in different parameter ranges. Perhaps the most important conclusion of these papers is that the very deep lattices (lattice potential $V$ larger than about $5E_R$) can be described very well by an extended fermionization model, in which most atoms localize in a Mott-insulator state with unit filling of the lattice, and the remaining atoms are free to move above the Mott state with a renormalized kinetic energy. Both the atoms in the Mott state and the remaining atoms are treated as effective non-interacting fermions whose dynamics are governed by a combination of the trap potential and appropriate kinetic energy terms.  These references also show, however, that there is a region of values of the ratio of interaction energy to kinetic energy (referred to as the ``intermediate region'' in \cite{ana2}) where the single-particle models, whether bosonic or fermonic, are inadequate to describe the dynamics of the Bose-Hubbard model, which is the main underlying theoretical tool for most of the studies described above.  This intermediate region, in the experiments of \cite{trey}, covers all the lattices studied with $V$ smaller than about $5E_R$, although there is some concern that for the shallowest lattices the tight-binding approximation leading to the Bose-Hubbard model itself may not be entirely accurate.

The present paper is an attempt to fill in this gap by presenting mean-field based calculations for the Bose-Hubbard dynamics in the ``intermediate region'' where the fraction of atoms having undergone the transition to the Mott insulator state is still negligible, and the system is mostly superfluid, yet the interaction energy cannot be neglected.  Our main calculational tool is time-dependent Hartree-Fock-Bogoliubov theory, and we show that this approach does indeed reproduce qualitatively many of the features of the damping observed in the experiments, although it generally underestimates its magnitude for any given lattice depth.  With the insights gained from these simulations, we develop a model for the damping that leads to a formula of the fluctuation-dissipation type, which relates the damping of the center of mass motion to the random density fluctuations in the noncondensate atoms that result from the dynamical instability.  We find that this formula leads to good agreement with the experimental data when parameters for the characteristic strength and correlation times of the fluctuations, obtained from the HFB calculations, are substituted in it. 

The basic ingredients of our model are:  (1) a large noncondensate fraction that arises as a direct consequence of the enhanced effective on-site interaction (due to the tight transverse confinement), of which a non-negligible part occupies high-momentum states and is therefore affected by the dynamical instabilities, and (2) the interaction between these noncondensate atoms and the condensate, which is modeled by exploiting a formal analogy with an external, random potential.  The latter is well known to determine localization
of the atomic wave-function in one-dimensional systems \cite{Anderson,Gavish}.
These ingredients (1) and (2) are introduced in Sections II and III, respectively. Section IV then presents the results of Hartree-Fock-Bogoliubov (HFB) calculations, which we use to estimate the parameters appearing in the damping formula.  Results from an alternative mean-field theory, based on the Gutzwiller ansatz, are presented in Section V.  Finally, Section VI is devoted to further discussions and conclusions.

\section{Hamiltonian and static (ground-state) results}

 The starting point for our
  theoretical treatment is a Hamiltonian of the ``tight binding'' or Bose-Hubbard form \cite{jaksch},
\begin{align}
\hat{H}=-J\sum_{<j,i>}\hat{a}_j^\dagger \hat{a}_{i}+\Omega \sum_j
j^2 \hat{n}_j + \frac U 2 \sum_j \hat{n}_j(\hat{n}_j-1)
\label{one}
\end{align}
In this expression, the sum $<i,j>$ is taken over nearest
 neighbors, $\hat{a}_j$($\hat{a}_j^{\dagger}$) are bosonic
field operators that annihilate (create) an atom at the lattice
site $j$, $\hat{n}_j=\hat{a}_j^{\dagger}\hat{a}_j$, and $\Omega=m
{\tilde{\omega}}_0^2 \lambda^2/(8 E_R)$ characterizes the
strength of the harmonic trap. The on-site interaction energy is
\begin{equation}
U=\frac{2a\hbar}{\sqrt{2 \pi}E_R}\left(\frac{\tilde{\omega}_x\tilde{\omega}_y\tilde{\omega}_z}{a_x
a_y a_z}\right)^{1/3}
\end{equation}
where $\tilde{\omega}_{x,y,z}$
 are the oscillation frequencies at individual lattice sites along the
three axes, obtained with a harmonic approximation expanding
around the minima of the potential wells,  $a_{x,y,z} =
\sqrt{\hbar/m\tilde{\omega}_{x,y,z}}$  are the effective harmonic
oscillator lengths and
 $a=5.31\times 10^{-9}$ m is the s-wave scattering length. For the
 experiment $\tilde{\omega}_x/2\pi = \tilde{\omega}_y/2\pi = 35$ kHz and  $\tilde{\omega}_z =  \sqrt{4 E_R^2 s}/\hbar$,
  where $s$ is the longitudinal periodic potential strength in units of the recoil energy, $s=V/E_R$.
  The ``hopping'' energy $J$ is
   $1/4$ of the  bandwidth, and obtained by the usual Mathieu function treatment (see \cite{ana2} for more details).  The definition (2) above implies that the parameters $U$,$J$ and $\Omega$ in (\ref{one}) are understood to be in units of $E_R$.  The summation indices in Eq.~(\ref{one}) range from $-M/2$ to $M/2$, where $M+1$ is the total number of wells (100--200 in our numerical calculations), and $j=0$ at the center of the trap.

\begin{figure}
\includegraphics[scale=0.86]{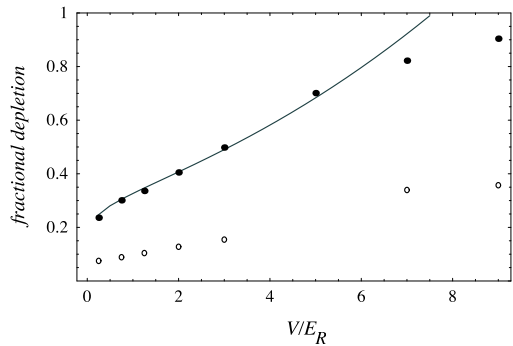}
\caption{\label{fig1}Black dots: fractional quantum depletion ($\tilde n/n = 1-n_c/n$) vs. lattice depth $V$. Open circles:  the fraction of $\tilde n$ with momenta $>\pi\hbar/\lambda$.  Solid line: Eq.~(\ref{two}) with $n=2.2$ (approximate atom density in the experiment at the center of the trap) and $\alpha=0.37$ (best fit).}
\end{figure}

We have used a numerical quantum Monte-Carlo \cite{prokofiev} method to derive the single-particle
density matrix for the Hamiltonian (\ref{one}) at very low temperature ($0.01J$, in our calculations), for a total number of atoms $N=80$.
 From it we obtain the quantum depletion shown in Figure 1.  The curve is a fit to the formula
\begin{equation}
1-\frac{n_c}{n} = \alpha\sqrt{\frac{U}{nJ}}
\label{two}
\end{equation}
which gives the depletion in the homogeneous case (that is, in the
absence of the harmonic trap, $\Omega=0$). $n_c$ is the density
of condensate atoms, $n$ the total density, and $\alpha$
a parameter that can be calculated from the Bogoliubov spectrum of
excitations \cite{anat,ana}.
 Since, in the trap potential, the spectrum is modified  and $n$ is not uniform,
$\alpha$ in the figure has been treated as an adjustable
parameter. We find that, even for very shallow lattices, some 20
to 40 percent of the atoms are not part of the condensate. The
figure also shows (open circles) the fraction of these
noncondensed atoms that have quasimomenta greater than $\hbar
\pi/\lambda$ (also calculated from the numerical single-particle
density matrix).  

In other studies, we have observed that the Mott insulator begins to form around $V=3E_R$ in this system, as characterized by a small decrease in the density fluctuations around the center of the trap that first becomes visible at this point.  Nonetheless, Figure 1 shows that the Bogoliubov result (\ref{two}) for $\tilde n$ remains approximately valid until around $V=5E_R$, which inspires us some confidence that the mean-field analysis that follows may be at least semi-quantitatively valid even for those very highly-depleted systems.

\section{A damping model}

In some recent work, one of us \cite{julio} has developed a formalism to describe the effect on matter waves of coherence-breaking processes such as random ``localizing'' events, momentum kicks, or perturbation by (time-dependent) random external potentials.  All these processes can be shown to lead to a damping of the center of mass motion of the system. 

Given the relatively large fraction of atoms in the dynamical instability region, calculated in the previous section, it seems, therefore, natural to consider their density fluctuations as providing a sort of random potential for the condensate to flow through, and to expect the damping to arise as a consequence of this.  The goal of this section is to make this picture plausible and quantitative, first by rederiving the damping due to an external random potential, then by showing how the Heisenberg equations of motion derived from the Hamiltonian (\ref{one}) can be cast into a similar form through a standard factorization ansatz, and, finally, deriving from all of this a damping formula.  Results from HFB calculations (to be discussed in much more detail in the next section) will also be introduced to establish the existence of the requisite random density fluctuations in these systems in the relevant regime.

In our tight-binding model, the center of mass position operator is 
\begin{equation}
\hat x_{cm}=\frac{\lambda}{2N}\sum_j j \hat n_j
\end{equation}
The commutator of $\hat x_{cm}$ with the
   Hamiltonian (\ref{one}) yields a center of mass velocity operator
\begin{equation}
\hat{v}=-\frac{iJ\lambda}{2N\hbar}\sum_j(\hat{a}_j^\dagger \hat{a}_{j+1} - \hat{a}_{j+1}^\dagger \hat{a}_j),
\end{equation}
which is essentially the same
    as the current operator in \cite{ana}. A further commutation yields the time derivative of
     $\hat{a}_j^\dagger \hat{a}_{j+1}$:
\begin{align}
i\hbar\frac{d}{dt}\hat{a}_j^\dagger \hat{a}_{j+1} = &(2j+1)\Omega \hat{a}_j^\dagger \hat{a}_{j+1} +
U\hat{a}_j^\dagger(\hat{n}_{j+1}-\hat{n}_j)\hat{a}_{j+1} \notag \\
&+ J(\hat{a}_{j-1}^\dagger \hat{a}_{j+1}-\hat{a}_j^\dagger
\hat{a}_{j+2}+\hat{n}_{j+1}-\hat{n}_j) \label{three}
\end{align}
with an analogous result for the derivative of
$\hat{a}_{j+1}^\dagger \hat{a}_j$.  In the ``intermediate region'' in which we are interested here, where the evolution of the system is not adequately described by single-particle models (bosonic or pseudo-fermionic), we expect the damping to arise from the
interaction term (proportional to $U$) in (\ref{three}).  

Before we get to work on that term, however, consider what would happen if one were to replace it in (1) by a random external potential, proportional to $\sum V_j \hat n_j$.  Eqs.~(4) and (5) would be unchanged, whereas Eq.~(\ref{three}) would become
\begin{align}
i\hbar\frac{d}{dt}\hat{a}_j^\dagger \hat{a}_{j+1} = &(2j+1)\Omega \hat{a}_j^\dagger \hat{a}_{j+1} +
(V_{j+1}-V_j)\hat{a}_j^\dagger \hat{a}_{j+1} \notag \\
&+ J(\hat{a}_{j-1}^\dagger \hat{a}_{j+1}-\hat{a}_j^\dagger
\hat{a}_{j+2}+\hat{n}_{j+1}-\hat{n}_j) 
 \label{6.1}
\end{align}
Now consider formally taking the ordinary quantum-mechanical expectation value of Eq.~(\ref{6.1}) and integrating it over an interval $(t-\Delta t,t)$, to get
\begin{align}
\av{\hat{a}_j^\dagger \hat{a}_{j+1}(t)} = &\av{\hat{a}_j^\dagger \hat{a}_{j+1}(t-\Delta t)} \notag\\
&-\frac{i}{\hbar}\int_{t-\Delta t}^t (V_{j+1}-V_j)(t')\av{\hat{a}_j^\dagger \hat{a}_{j+1}(t')} dt' +\ldots
\label{6.2}
\end{align}
where $\ldots$ represents terms that do not contain $V_j$'s.  Substituting (\ref{6.2}) back into the (expectation value of the) second term on the right-hand side of (\ref{6.1}), one obtains two kinds of terms: some linear in the $V_j$, and some quadratic in $V_j$.  The linear ones involve products of a $V_j$ at the time $t$ and field operators at an earlier time, and we may assume that they vanish in an ensemble average over different realizations of the random process $V_j$, provided it has a sufficiently short correlation time.  The ensemble average of the quadratic terms, on the other hand, yields
\begin{align}
&-\frac{i}{\hbar}\int_{t-\Delta t}^t \av{(V_{j+1}-V_j)(t)(V_{j+1}-V_j)(t')}\av{\hat{a}_j^\dagger \hat{a}_{j+1}(t')} dt'  \notag\\
&\simeq -\frac{i}{\hbar}\tau_c \av{(V_{j+1}-V_j)^2} \av{\hat{a}_j^\dagger \hat{a}_{j+1}(t)}
\label{6.3}
\end{align}
where $\tau_c$ is the characteristic correlation time for the randomly-fluctuating potential, $\av{(V_{j+1}-V_j)^2}$ is the average (squared) strength of the fluctuations, and it has been assumed that $\av{\hat{a}_j^\dagger \hat{a}_{j+1}}$ (essentially, the velocity of the system) does not change appreciably over the time scale of $\tau_c$.  (This is, basically, the Markov approximation.)  The result, since the left-hand side of (7) is multiplied by $i\hbar$, is clearly a damping term for $\av{\hat{a}_j^\dagger \hat{a}_{j+1}}$, or, by (5), for the on-site velocity $v_j\equiv i\av{\hat{a}_j^\dagger \hat{a}_{j+1} - \hat{ a}_{j+1}^\dagger \hat{ a}_j}$:
\begin{equation}
\frac{d v_j}{dt} = -2\gamma_j v_j +\ldots
\end{equation}
with
\begin{equation}
\gamma_j = \frac{1}{2\hbar^2}\tau_c \av{(V_{j+1}-V_j)^2} 
\end{equation}

The question now is whether it is possible to extract, from the interaction term in (6) something that looks like an ``external'' random potential, as in Eq.~(7).  That this is in fact possible follows if one replaces the bosonic field operators $\hat a_j$ by $\hat a_j = z_j+\hat\delta_j$, where $z_j$ is a c-number equal to $\av{\hat a_j}$ (the local mean-field), and $\hat\delta_j$ a zero-average operator.  Substituting in the interaction term in (6), we get
\begin{widetext}
\begin{align}
\hat{a}_j^\dagger(\hat{n}_{j+1}-\hat{n}_j)\hat{a}_{j+1} = &(|z_{j+1}|^2 - |z_j|^2)\hat{a}_j^\dagger \hat{a}_{j+1} \notag \\
&+\hat{a}_j^\dagger\left(z_{j+1}^\ast \hat\delta_{j+1} +  z_{j+1}\hat\delta_{j+1}^\dagger + \hat\delta_{j+1}^\dagger \hat\delta_{j+1}-z_j^\ast \hat\delta_j -  z_j\hat\delta_j^\dagger - \hat\delta_j^\dagger \hat\delta_j\right)\hat{a}_{j+1}
\label{e40}
\end{align}
The first term on the right-hand side of (\ref{e40}) is a ``deterministic'' term which can be combined with the first term on the right-hand side of (\ref{6.1}); indeed, it is the combination of these two terms that yields the Thomas-Fermi profile in the ground state when the kinetic energy term (the $J$ term) in (\ref{6.1}) is negligible.  The second term in (\ref{e40}), on the other hand, is where we expect the main ``noise'' to arise.  To determine its contribution to the equation of motion for the expectation value $\av{\hat{a}_j^\dagger \hat{a}_{j+1}}$, we express the remaining $\hat a_j$ operators in terms of $\hat\delta_j$, assume expectation values of the form $\av{(\hat\delta^\dagger)^p \hat\delta^q}$ vanish unless $p=q$, and factor terms such as $\av{\hat\delta_j^\dagger \hat\delta_j^\dagger\hat\delta_j\hat\delta_{j+1}}$ in a standard way, as $2\av{\hat\delta_j^\dagger \hat\delta_j}\av{\hat\delta_j^\dagger\hat\delta_{j+1}}$.  The result is
\begin{align}
\Bigl\langle\hat{a}_j^\dagger\Bigl(z_{j+1}^\ast \hat\delta_{j+1} +  z_{j+1}\hat\delta_{j+1}^\dagger &+ \hat\delta_{j+1}^\dagger \hat\delta_{j+1}-z_j^\ast \hat\delta_j -  z_j\hat\delta_j^\dagger - \hat\delta_j^\dagger \hat\delta_j \Bigr)\hat{a}_{j+1} \Bigr\rangle  \notag \\
&=2\left(\left\langle \hat\delta^\dagger_{j+1}\hat\delta_{j+1}\right\rangle-\left\langle \hat\delta^\dagger_{j}\hat\delta_{j}\right\rangle\right)\left(z_j^\ast z_{j+1}+\av{\hat\delta_j^\dagger \hat\delta_{j+1}}\right) + (|z_{j+1}|^2 - |z_j|^2)\left\langle\hat{\delta}_j^\dagger \hat{\delta}_{j+1}\right\rangle\notag\\
&=2(\tilde n_{j+1}-\tilde n_j)\av{\hat{a}_j^\dagger \hat{a}_{j+1}}+(|z_{j+1}|^2 - |z_j|^2)\left\langle\hat{\delta}_j^\dagger \hat{\delta}_{j+1}\right\rangle
\label{e41}
\end{align}
\end{widetext}
where the noncondensate density $\tilde n_j \equiv \av{\hat n_j} -|z_j|^2$ has been introduced.  The second term on the right-hand side of (\ref{e41}) appears to be a small noise-induced contribution to the deterministic part of (\ref{e40}).  The first term has the desired form.  We can then replace the expectation value of the interaction term in (6) by a deterministic term, with which we shall not be concerned any more, and a noise-like term
\begin{equation}
2U(\tilde n_{j+1}-\tilde n_j)\av{\hat{a}_j^\dagger \hat{a}_{j+1}}
\end{equation}
If the evolution of the $\tilde n_j$ is sufficiently chaotic, one could imagine integrating the Heisenberg equations many times for very slightly different initial conditions and obtaining each time a different realization of the ``random process'' $\tilde n_j$.  Then, if the Markovian condition holds, one can follow the same steps as for the external potential $V_j$ in Eqs.~(7)--(11) above and, by identifying $V_j$ with $2U\tilde n_j$, conclude that, on average, a damping
\begin{equation}
\gamma_j = \frac{2U^2}{\hbar^2}\tau_c \av{(\tilde n_{j+1}-\tilde n_j)^2}  \equiv \frac{2U^2}{\hbar^2}\tau_c \av{f_j^2} 
\label{gammaj}
\end{equation}
will be observed in this system, for the on-site velocity $v_j$.  (For conciseness, we have introduced the notation $f_j\equiv \tilde n_{j+1}-\tilde n_j$).  The overall damping of the center of mass motion could be estimated by taking a weighted average of the $\gamma_j$ (although, if the $\gamma_j$ are very different from site to site, the assumption of a single damping constant for the center of mass motion may not be a very good approximation).

Our model is, therefore, that the condensate atoms are slowed down as they attempt to move through
a randomly fluctuating effective potential created (through the interaction term) by the noncondensate atoms, as the latter are ``shaken'' out of equilibrium by the displacement of the trap.  It may be worthwhile, at this point, to go over and attempt to justify the various assumptions that have been made.  

The interpretation of $\tilde n_{j+1}-\tilde n_j$ as an essentially random variable appears justified from time-dependent HFB calculations (about which much more will be said in the next section) such as the one illustrated in Fig. 2 for a lattice of depth $V= 1E_R$: the top part shows the time trace of $\tilde n_1-\tilde n_0$, and the bottom figure the logarithm of the absolute value of its (time-)autocorrelation function, with a linear fit showing an approximately exponentially decaying envelope.   It is, however, not quite as clear whether the Markov approximation is valid: after all, $\tilde n_j$ is not an external field, but one of the system's dynamical variables, and it certainly must develop correlations and become entangled with other dynamical variables as the system evolves.  Still, we take this as the simplest approximation, and note that, as will be seen in the next section, over the range considered, the time scale $\tau_c$ for the decay of correlations in  $\tilde n_{j+1}-\tilde n_j$ (which is, essentially, $\hbar/J$) is indeed well-separated from the time scale of the damping of the center of mass oscillations.  
 
\begin{figure}
\includegraphics[scale=0.9]{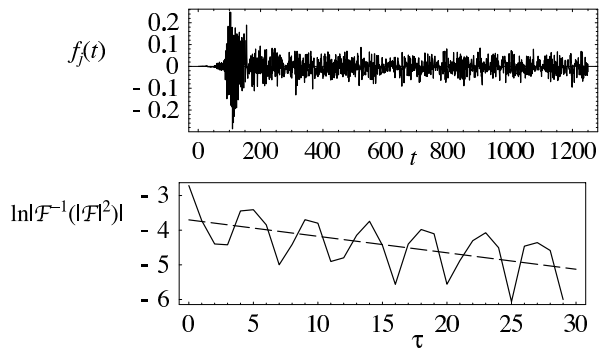}
\caption{\label{fig2} (a) $\tilde n_{j+1} - \tilde n_j$ for $V=1
E_R$ and $j=0$; (b) Fit (dashed line) to $\ln|  {\cal F}^{-1}
\left[ |{\cal F}[f_j(t)]|^2 \right]|$, for the first 30 time
steps, for the $f_j$ shown in (a).  The slope of this line is
taken to be $1/\tau_c$, for this particular value of $j$.  Time is
in units of $\hbar/E_R$ in both cases.}
\end{figure}

Besides the above approximations, we have neglected ``anomalous averages'' such as, e.g., $\av{\hat\delta_j\hat\delta_j}$, and we have used a standard ``bosonic'' ansatz to factor expectation values of products of four operators into expectation values of products of two operators.  We do this in the spirit of all mean-field theories; namely, as something to try and see how it works.  We certainly do not expect it to be a good approximation once (extended) fermionization becomes important.

Note that if, instead of using the bosonic asatz, we had taken the simplest approach of factoring $\av{\hat{a}_j^\dagger(\hat{n}_{j+1}-\hat{n}_j)\hat{a}_{j+1}}$ as $\simeq\av{\hat{a}_j^\dagger\hat{a}_{j+1}}\av{\hat{n}_{j+1}-\hat{n}_j}$ (and then separated out the condensate part from $\av{\hat{n}_{j+1}-\hat{n}_j}$), the resulting ``noise'' term would have differed from (14) by a factor of 2, and hence the damping formula (\ref{gammaj}) would have been four times smaller.  This may be a reasonable estimate of the possible error involved in our factorization assumptions.

Equation (\ref{gammaj}) does not, by itself, tell us what the actual damping is; for that, one needs to know the parameters characterizing the strength of the noise, $\av{f_j^2}$, and its characteristic correlation time $\tau_c$.   A possible way to obtain a very rough order-of-magnitude estimate for these quantities has been sketched in \cite{julio}, by rewriting Eq.~(\ref{gammaj}) in terms of the discrete Fourier transform (momentum components) of the $\tilde n_j$ (the order of magnitude of which can be estimated from Fig.~1), and assuming that $\tau_c$ should be of the order of magnitude of $\hbar/J$, since this is the ``hopping rate,'' and that is the time scale over which one would expect local density fluctuations to decay.  This simple approach does indeed yield the order of magnitude of the experimentally-observed damping.  

As we shall show below, using values for $\av{f_j^2}$ and  $\tau_c$ derived from HFB calculations in the formula (\ref{gammaj}) does lead to very good agreement with the experimentally-observed damping. 

\section{HFB calculations}

In this section we report on the results of calculations using the  time dependent
Hartree-Fock-Bogoliubov (HFB) approximation
\cite{anaHFB,Griffin1}. The starting point of this approximation
is the  Heisenberg equation of motion for the field operator:
\begin{equation}
 i\hbar\frac{d}{dt}\hat{a}_j =(\hat{K}+ \Omega j^2+
U\hat{a}_j^\dagger\hat{a}_{j})\hat{a}_{j}, \label{ana1}
\end{equation}
with $\hat{K}$ the tight binding kinetic energy operator:
$\hat{K}A_j=-J(A_{j+1}+A_{j-1})$ (here $A_j$ can be any function or operator defined at the point $j$), and $\Omega j^2$ the external confining
potential, which is quadratic in our system. By expressing the
field operator, as before, as $\hat{a}_j =z_j+\hat{\delta}_j$,
replacing this ansatz in Eq. (\ref{ana1}), and treating the cubic
term in a self-consistent mean field approximation, coupled
equations of motion  for the condensate, $z_j$, and the
fluctuating field $\hat{\delta}_j$ can be obtained:
\begin{align}
i\hbar\frac{d}{dt}z_j &= (\hat{K}+\Omega j^2+
U(|z_j|^2+2\tilde{n}_j))z_j+U\tilde{m}_jz_j^*,\\
i\hbar\frac{d}{dt}\hat{\delta}_j &= (\hat{K}+ \Omega j^2+
2Un_{j})\hat{\delta}_j+Um_{j}\hat{\delta}_{j}^\dagger,\label{ana3}
\end{align}
\noindent here $n_j=\langle\hat{a}_j^\dagger\hat{a}_j\rangle$,
$\tilde{n}_j=\langle\hat{\delta}_j^\dagger\hat{\delta}_j\rangle$,
$m_j=\langle\hat{a}_j\hat{a}_j\rangle$ and
$\tilde{m}_j=\langle\hat{\delta}_j\hat{\delta}_j\rangle $.
 By using a Bogoliubov transformation that expresses the
operators $\hat\delta_j$ in terms of quasiparticle creation and
annihilation operators $\hat{\alpha}_k,\hat{\alpha}_k^\dagger$ and
amplitudes $\{u_j^k(t)\},\{v_j^k(t)\}$, as
$\hat\delta_j=\sum_k(u_j^k(t)\hat{\alpha}_k-{v_j^k}^*(t)\hat{\alpha}_k^\dagger)$,
one  obtains  equations of motion for the amplitudes $\{z_j\}$,
$\{u_j^k(t)\}$ and  $\{v_j^k(t)\}$ known as HFB equations. They
 describe the coupled dynamics of condensate and noncondensate
atoms and  conserve particle number and energy. As coupled, nonlinear equations, they have the potential to describe a wide range of dynamics, including deterministic chaos.

A difficulty with the time-independent HFB equations is that they violate the
Hugenholtz-Pines theorem and yield an initial ground state with a
depletion that is too small \cite{anat}, when compared to the
exact numerical results in Fig.~1.  We have therefore used the
Popov approximation \cite{Griffin1} (which ignores the anomalous
terms $\tilde{m}_j$) to calculate the ground state of the
undisplaced trap, but then, after displacing the trap, we
propagate in time using the full HFB equations (without the Popov
approximation), because propagating in time with the Popov
approximation does not conserve particle number or energy.  Due to
this mixing of approximations (in a sense, we are starting from
the ``wrong'' initial state), as well as to the intrinsic
limitations of the HFB approximation, our HFB results must be
taken with some caution. Nonetheless, one may gain at least some
qualitative insights from them, as illustrated, for instance, in
Fig.~3, which shows how the noncondensate atoms relax rather
rapidly (in agreement with the expectation of strongly inhibited
transport for the high-momentum states), and with a substantial
amount of noise.

\begin{figure}
\includegraphics[scale=0.6]{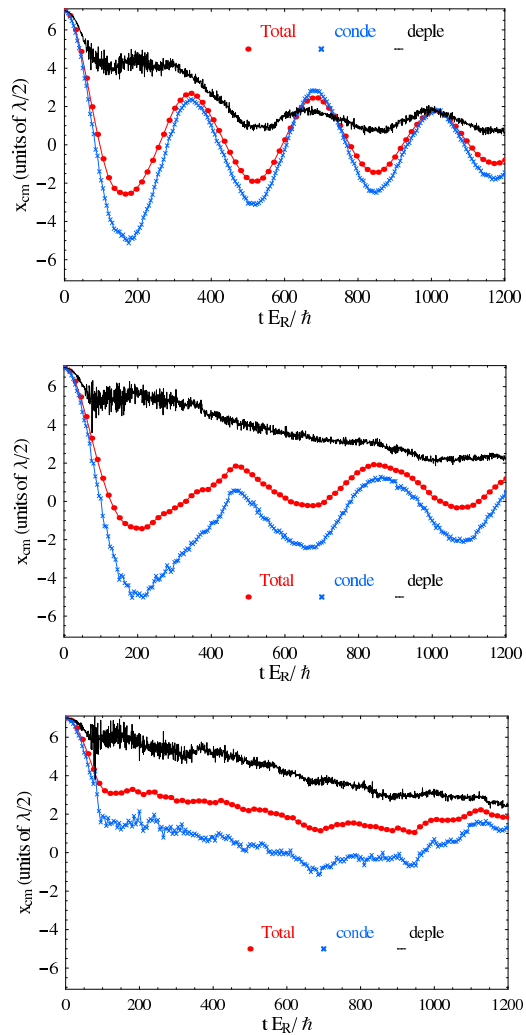}
\caption{\label{fig3} Result of the direct numerical integration of the HFB equations for, from top to bottom, $V=3E_R$, $V=4E_R$, and $V=5E_R$.  In all cases the position of the center of mass of the noncondensate atoms is given by the noisier top trace, that of the condensate by the lighter lower trace (small crosses), and the total is given by the line drawn with the circles. Time is in units of $\hbar/E_R$ and position
 is in units of the lattice spacing.}
\end{figure}

Our HFB calculations do exhibit damped center of mass oscillations for all the values of $V$ in the experiment, and for sufficiently deep lattices (about $V>4E_R$ in our calculations) they even exhibit the overdamped relaxation seen in the experiments (i.e., the value of the damping exceeds the oscillation frequency), although in the experiments this transition to overdamped motion was seen already for shallower lattices, between  $V=2E_R$ and $V=3E_R$.   Quantitatively speaking, the HFB results do predict, in general, a smaller damping than is seen experimentally for any given lattice depth $V$, and also, even with the Popov approximation, a smaller ground-state depletion than the Monte-Carlo calculations in Fig.~1.  This lack of precise quantitative agreement is not terribly surprising, given the fact that for all of these systems the depletion of the condensate is really not very small when compared to the mean field density; hence neglecting higher powers of the $\hat\delta_j$ operators cannot be very accurate.  The qualitative agreement, however, suggests that the HFB approximation does retain all the physical ingredients needed to predict the kind of damped oscillations seen in the experiments in this regime, 

With all of the above in mind,  we have attempted to use the results of the HFB calculations to estimate the quantities $\av{f_j^2}$ and $\tau_c$ in the damping formula \ref{gammaj}, in the following manner. First, we generate time series for $\tilde
n_j(t)$ for all $j$ and for a relatively large number of
oscillation periods, and we simply average all these values to
estimate $\av{f_j^2(t)}$.  We also calculate the Fourier transform
$\tilde f_j(\omega) \equiv {\cal F}(f_j)$ of each time series, and
then calculate the inverse Fourier transform, ${\cal F}^{-1}$, of
the power spectrum $|\tilde f_j(\omega)|^2$; by the convolution theorem of Fourier transforms, this should equal the autocorrelation of $f_j(t)$.  We therefore
estimate a correlation time by fitting an exponential to the decay
of the absolute value of ${\cal F}^{-1}(|\tilde
f_j|^2)$,  for relatively short times (of the order of
$30\hbar/E_R$).  (Representative results are shown in Fig.~2(b).)  We
obtain in this way a (generally different) value of $\tau_c$ and
$\av{f_j^2(t)}$ for every lattice site $j$.  The final estimate of
the overall damping $\Gamma$ is obtained by taking a weighted
average of all the $\gamma_j$, using the equilibrium density as
the weighting function. This results in the gray dots in Fig.~4,
which are to be compared to the experimental data shown as the
black dots in the same figure.

\begin{figure}
\includegraphics[scale=0.9]{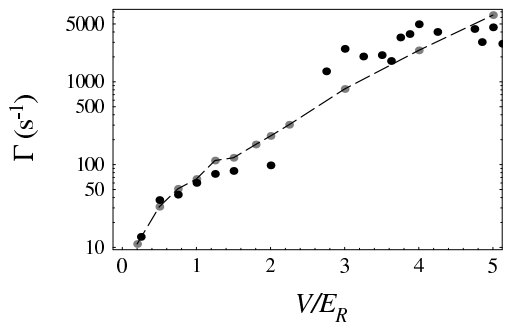}
\caption{\label{fig4} Gray dots (connected by dashed line):
the value of $\Gamma$ calculated from Eq.~(3), using the time series
 that result from integrating the HFB equations.  Black dots: experimental data. }
\end{figure}

We are faced with the somewhat paradoxical result that, while the HFB calculations generally underestimate the damping, the formula (\ref{gammaj}), using HFB values, agrees quite well with the experiments and even appears to overestimate the damping in places (such as around $V=2E_R$).  

We do not have, in principle, a reason to doubt the relative accuracy of the HFB estimate of the fluctuations' correlation time $\tau_c$ (which does turn out to be between $2\hbar/J$ and $3\hbar/J$ for the values of $V$ considered).  On the other hand, the fact, pointed out above, that the HFB calculations predict a noncondensate density lower than the true one, suggests that the HFB estimate of $\av{f_j^2(t)}$ may be proportionately low as well.  If this is the case, it would indicate that the formula (\ref{gammaj}) generally overestimates the damping, perhaps because of the assumption of totally uncorrelated condensate and noncondensate fluctuations that goes into its derivation.  The agreement with the experiment shown in Fig.~4 would then appear to be more precise that it is actually supposed to be.  Nonetheless, generally speaking, the physical picture invoked in the derivation of the damping formula appears to be correct, even if oversimplified in some details (e.g., validity of the Markov approximation).   

\section{Results from Gutzwiller-ansatz calculations}

It is well-known that an alternative to the HFB calculations is provided by a mean-field theory based on the Gutzwiller ansatz.  While in the HFB method the interaction term in (1) is treated approximately, and the kinetic energy term is treated exactly, yielding a theory best suited for weakly interacting superfluids, the Gutzwiller ansatz is equivalent to treating the interaction term in (1) exactly and approximating the kinetic energy term as follows
\begin{align}
\hat a_{j+1}^\dagger \hat a_j &= \Bigl[\av{\hat a_{j+1}^\dagger} + \bigl(\hat a_{j+1}^\dagger - \av{\hat a_{j+1}^\dagger}\bigr)\Bigr]\Bigl[\av{\hat a_j} + (\hat a_j - \av{\hat a_j})\Bigr] \notag \\
&\simeq \av{\hat a_{j+1}}^\ast \hat a_j + \hat a_{j+1}^\dagger \av{\hat a_j} - \av{\hat a_{j+1}}^\ast \av{\hat a_j}
\label{e46}
\end{align}
As a result of this, the Gutzwiller ansatz works best for very strongly interacting systems, and it is in fact capable of describing qualitatively the Mott transition \cite{jaksch2}, which the HFB method cannot do.

In Eq.~(\ref{e46}), the mean field $\av{\hat a_j}$ is obtained self-consistently by diagonalization of the resultant effective Hamiltonian, to which a chemical potential term $-\mu\sum n_j$ is added in order to get the desired average number of particles.  Once the initial state has been calculated, the relevant equations of motion are as given, for instance, in \cite{altman,jaksch2}.

What we find from the Gutzwiller approach is that the predicted depletion for the ground state is substantially lower than the one calculated numerically in Fig.~1, for all except the deepest lattices, and accordingly no appreciable damping is seen, for the experimental parameters, until $V=4E_R$ or so. For $V=3E_R$ a displacement of the trap potential by 6 lattice sites fails to give any visible damping, but a displacement of 8 lattice sites does give substantial damping, as shown in Figure 5(a).  

\begin{figure}
\includegraphics[scale=0.65]{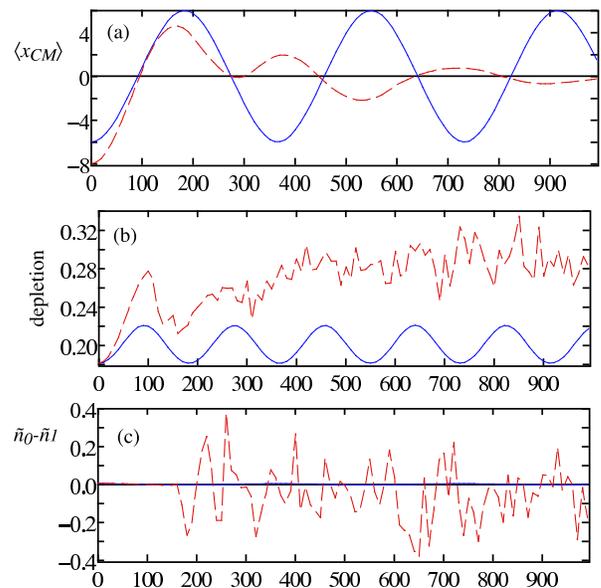}
\caption{\label{fig5} Results of Gutzwiller calculations for $V=3E_R$ and initial displacements $d=6$ (solid line) and $d=8$ (dashed line). (See text for details.)}
\end{figure}

The other graphs in Figure 5, also for $V=3E_R$, highlight other interesting features of this transition from undamped to damped motion, which are in rough agreement with our prior expectations.  Figure 5(b) shows the fractional depletion as a function of time for the two displacements $d=6$ (solid) and $d=8$ (dashed).  Although the initial depletion is the same (the ground state value), the time evolution leads to a depletion that, in the case of regular motion, is largest at the times when the condensate is moving faster.  In the case $d=8$ one can see the depletion initially growing as the speed (the slope of the corresponding curve in Fig.~5(a)) increases, and eventually becoming rather large, after which damped motion follows.  Note, for reference, that the Monte-Carlo prediction from Fig.~1 for this case would be a ground state depletion of about $0.5$.  On the other hand, the smallest depletion calculated in Fig.~1, for $V=0.25 E_R$, is $0.24$, which here would appear to be just large enough to result in damped motion.

Figure 5(c) shows the time dependence of $\tilde n_0 - \tilde n_1$ (the subscript ``0'' refers to the center of the trap).  The regular case, for $d=6$, is a solid line invisible on the scale of the figure ($<0.005$ in magnitude).  Again, the damping appears to be strongly correlated with the site-to-site density fluctuations.  

As we did for the HFB calculation, we can extract values for $\av{f_j^2}$ and $\tau_c$ from the time series obtained by the Gutzwiller approach, and substitute them in Eq.~(\ref{gammaj}).  The results, for $V/E_R = 4, 5$, are actually quite close to those obtained from the HFB calculation.  For smaller $V$, of course, the calculation would not make sense, since the time dependence of $\tilde n_j - \tilde n_{j+1}$ predicted by the Gutzwiller ansatz in this region is always regular, rather than noise-like.

\section{discussion and conclusions}

To recapitulate, then, we believe that the damping in this ``intermediate'' regime can be explained as arising
from the large depletion due to the tight transverse confinement,
which leads to the population of high-momentum states in the
non-quadratic part of the lattice dispersion curve. The condensate
atoms' motion is then damped through their interaction with the
random field created by these noncondensate atoms when their
equilibrium state is perturbed. The dramatic growth of $\Gamma$
with $V$ illustrated in Fig.~4 arises from several causes:
 first, the depletion increases with lattice depth, as shown by Fig.~1 and Eq.~(\ref{two});
  second, the interaction $U$ itself increases, albeit weakly (as $V^{1/4}$); third,
  the tunneling rate $J$ decreases, and the correlation time $\tau_c \sim \hbar/J$ in (\ref{gammaj})
  increases accordingly (the ``damping medium'' becomes more ``sluggish'').  

The main limitations of the formula (15) have been pointed out when it
was derived. Since it only accounts for the damping induced by the
interactions, it vanishes in the $U \to 0$ limit, even though, as we mentioned in the Introduction, a noninteracting bosonic
gas exhibits a sort of damping in a lattice, associated with the non-
harmonic nature of the total potential (see, e.g., \cite{ana2}).  At the other limit point, $U \to \infty$, Eq.~(15) predicts an infinite
damping, which is clearly also not correct. The reason is that
Eq.~(15) is based on a self-consistent factorization approximation
that, strictly speaking, is only valid in the weakly interacting
limit. 

We note that, in these regimes were Eq.~(15) does not apply, previous
studies \cite{Ruuska,Ott,ruostekoski,guido,Rigol}
have shown that treatments
based on single-particle solutions can provide a very accurate
description of the damping.  On the other hand, in the complex intermediate regime where
it is not possible to use the simplicity of the single-particle
solutions, we have shown that Eq.~(15) does manage to describe the
damping. Moreover, it connects in a simple way the damping rate to
physical parameters and therefore allows a clearer understanding of the
physics responsible for the dissipative dynamics exhibited by  1D
lattice systems in this regime.

\section{Acknowledgements}

J. Gea-Banacloche is grateful to NIST for hospitality, and ARO for partial support.  All the authors acknowledge many helpful discussions with James (Trey) Porto and W. D. Phillips, and their team.


\begin{thebibliography}
\bibitem{}

\bibitem{Ruuska} V. Ruuska and P. T\"{o}rm\"{a}, New Journal of Physics {\bf 6},59 (2004).
\bibitem{Pezze} L. Pezz\`{e}, L. Pitaevskii, A. Smerzi and S. Stringari, Phys. Rev. Lett. \textbf{93}, 120401 (2004).  
\bibitem{Ott} H. Ott, E. de Mirandes, F. Ferlaino, G. Roati, V. T\"urck, G. Modugno and M. Inguscio,  cond-mat/0404201. 
\bibitem{Rigol} M. Rigol and A. Muramatsu, Phys. Rev. A \textbf{70}, 043627 (2004). 
\bibitem{Rigol3}  M. Rigol, V. Rousseau, R. T. Scalettar and R. R. P. Singh, cond-mat/0503302 (2005). 
\bibitem{ana2} Ana Maria Rey, Guido Pupillo, Charles W. Clark, and Carl J. Williams, cond-mat/0503477, PRA accepted. 
\bibitem{wu} B. Wu, R. B. Diener, and Q. Niu, Phys. Rev. A {\bf 65}, 025601 (2002), and references therein. 
\bibitem{burger} S. Burger, F. S. Cataliotti, C. Fort, F. Minardi, M. Inguscio, M. L. Chiofalo and M. P. Tosi, Phys. Rev. Lett. {\bf 86}, 4447 (2001). 
\bibitem{Smerzi} A. Smerzi, A. Trombettoni, P. G. Kevrekidis and A. R. Bishop, Phys. Rev. Lett. \textbf{89}, 170402 (2002); 
\bibitem{trey} C. D. Fertig, K. M. O'Hara, J. H. Huckans, S. L. Rolston, W. D. Phillips, and J. V. Porto, Phys. Rev. Lett. {\bf 94}, 120403 (2005). 
\bibitem{esslinger} Thilo St\"oferle, Henning Moritz, Christian Schori, Michael K\"ohl, and Tilman Esslinger, Phys. Rev. Lett. {\bf 92}, 130403 (2004). 
\bibitem{Morsch} O. Morsch, J. H. M\"uller, M. Cristiani, D. Ciampini, and E. Arimondo, Phys. Rev. Lett. {\bf 87}, 140402 (2001). 
\bibitem{Cataliotti} F. S. Cataliotti, S. Burger, C. Fort, P. Maddaloni, F. Minardi, A. Trombettoni, A. Smerzi and M. Inguscio, Science {\bf 293}, 843 (2001). 
\bibitem{polkovnikov} A. Polkovnikov and D.-W.Wang, Phys. Rev. Lett. {\bf 93}, 070401 (2004).
\bibitem{altman} E. Altman, A. Polkovnikov, E. Demler, B. Halperin, and M. D. Lukin, Phys. Rev. Lett. {\bf 95}, 020402 (2005).
\bibitem{polkovnikov2} A. Polkovnikov, E. Altman, E. Demler, B. Halperin, and M. D. Lukin, Phys. Rev. A {\bf 71}, 063613 (2005).
\bibitem{ruostekoski} J. Ruostekoski and L. Isella, cond-mat/0504026.
\bibitem{guido} Guido Pupillo, Ana Maria Rey, Carl J. Williams, and Charles W. Clark, cond-mat/0505325. 
\bibitem{Anderson} P.W. Anderson, Phys. Rev. {\bf 109}, 1492 (1958).
\bibitem{Gavish} U. Gavish and Y. Castin, Phys. Rev. Lett. {\bf 95}, 020401 (2005).
\bibitem{jaksch} D. Jaksch, C. Bruder, J. I. Cirac, C. W. Gardiner, and P. Zoller, Phys. Rev. Lett. {\bf 81}, 3108 (1998).
\bibitem{prokofiev} N.V. Prokof'ev,  B. V. Svistunov, and I. S. Tupitsyn, Phys. Lett. A {\bf 238}, 253 (1998); Sov. Phys. JETP 87, 310 (1998).
\bibitem{anat} A. M. Rey, D. Phil. Thesis, University  of Maryland at College Park (2004).
\bibitem{ana} A. M. Rey, K. Burnett, R. Roth, M. Edwards, C. J. Williams and C. W. Clark, J. Phys. B: At. Mol. Opt. Phys. {\bf 36}, 825 (2003). 
\bibitem{julio} J. Gea-Banacloche, quant-ph/0508134.
\bibitem{griffin} J. E. Williams and A. Griffin, Phys. Rev. A {\bf 63}, 023612 (2001).
\bibitem{anaHFB} Ana Maria Rey, B. L. Hu, E. Calzetta, A. Roura and C. W. Clark Phys. Rev. A \textbf{69}, 033610 (2004).
\bibitem{Griffin1} A. Griffin, Phys. Rev. B \textbf{53}, 9341 (1995).
\bibitem{Giorgini} S. Giorgini, Phys. Rev. A \textbf{57} 2949 (1997); S. Giorgini, Phys. Rev. A \textbf{61}, 063615 (2000).
\bibitem{jaksch2} D. Jaksch, V. Venturi, J. I. Cirac, C. J. Williams, and P. Zoller, Phys. Rev. Lett. {\bf 89}, 040402 (2002).


\end{thebibliography}
\end{document}